\documentclass[aps,prl,twocolumn,groupedaddress]{revtex4}

\usepackage{amsmath}
\usepackage{graphicx}

\begin{document}

\title{Tunneling through an Aharonov-Bohm ring -- effects of
dephasing by electron-electron interactions}

\author{R. \v{Z}itko}
\author{J. Bon\v{c}a}

\affiliation{FMF, University of Ljubljana, 
and J. Stefan Institute, Ljubljana, Slovenia}

\date{\today}

\begin{abstract}
We develop a non-perturbative numerical method to study a single
electron tunneling through an Aharonov-Bohm ring in the presence of
bound, interacting electrons.  Inelastic
processes and spin-flip scattering are properly taken into
account. We show that
electron-electron interactions described by the Hubbard Hamiltonian
lead to strong dephasing and we obtain high transmission probability
at $\Phi=\pi$ even at small interaction strength. Depending on the
many-electron state on the ring, dephasing can occur in elastic or
inelastic channels with or without changing the spin of the
scattering electron.
\end{abstract}


\maketitle

Quantum interference can be studied in mesoscopic systems 
where the wave nature of electrons plays an important role.  
Particularly noteworthy are
studies of the Aharonov-Bohm (AB) oscillations in mesoscopic rings
\cite{webb,timp}.
Inelastic scattering of
electrons is believed to be predominantly responsible for the loss of
the phase coherence in such experiments. 
When an electron interacts with optical phonons, the dephasing only
occurs through inelastic processes \cite{bonca3}.
At low temperatures the phonon degrees of freedom 
freeze out, therefore other mechanisms for dephasing, like
magnetic impurity scattering or zero-point fluctuations of the
electromagnetic environment \cite{mohanty1,mohanty2} have been
proposed. It is nevertheless believed, that at low temperatures
the electron-electron (e-e) interaction is a dominant mechanism for
dephasing \cite{mesoscopic}.

The AB geometries have been theoretically studied predominantly 
by  self-consistent
mean-field approximations that break down for degenerate levels, which
physically happens at very low temperatures
\cite{phasechange,phasechange2,reflection}.
They do not describe transitions 
in which the symmetry of the wavefunction of bound
electrons changes and they are thus inadequate to study decoherence.
Renormalisation group techniques have
been applied to AB systems \cite{rg}, where calculations were limited
to spinless interacting quantum dot with two levels coupled to
reservoirs. Particular attention was devoted to the appearance of Kondo
physics induced by changing magnetic flux, however due to limitations
to spinless fermions, no spin-flip induced decoherence has been
investigated by this method.

To shed some new light on the problem of decoherence, 
there is obviously a demand for a
capable method, that would treat the problem of the scattering of an
electron through a limited region where e-e interactions would be exactly
taken into account.  In this Letter we propose a method
that treats e-e interactions by direct diagonalisation using iterative
(Lanczos) technique.  
The method takes into account spin-flip processes, 
so it can also be used in calculations of spin-polarized transport \cite{ab}.

We apply the method to study a single-electron
transmission through a ring with e-e interactions described 
by a Hubbard Hamiltonian. We show that
dephasing can occur either by a) inelastic processes where the
tunneling electron excites bound electrons on the ring or by b)
elastic processes, where the tunneling electron changes the symmetry
of the degenerate many-electron wavefunction.
No exchange of energy is required in the
latter case \cite{phase,model,ab,interferometry}:
dephasing occurs because the tunneling electron leaves a trace on its
``environment'', which consists of bound electrons. In either elastic 
or inelastic case the dephasing can occur with or without
the spin-flip of the scattering electron. 

\newcommand{\bra}[1]{\langle {#1} |}
\newcommand{\ket}[1]{| {#1} \rangle}
\newcommand{\braket}[2]{\langle {#1} | {#2} \rangle}
\newcommand{\upa}{\uparrow}
\newcommand{\dna}{\downarrow}
\newcommand{\cc}[2]{c_{#1,#2}}
\newcommand{\cp}[2]{{c_{#1,#2}}^{\dag}}
\newcommand{\cpm}[2]{{a_{#1,#2}}^\dag}
\newcommand{\cpn}[2]{{b_{#1,#2}}^\dag}
\newcommand{\cm}[2]{{a_{#1,#2}}}
\newcommand{\cn}[2]{{b_{#1,#2}}}
\newcommand{\rng}{{\text{ring}}}
\newcommand{\lll}{{\text{L}}}
\newcommand{\rr}{{\text{R}}}

\begin{figure}
\includegraphics[height=3.5cm]{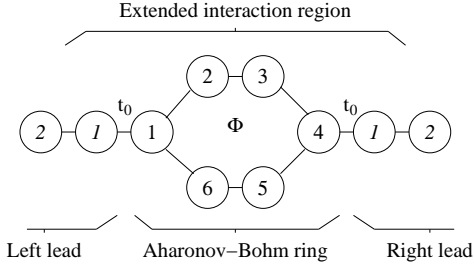}
\caption{Aharonov-Bohm ring. Magnetic flux $\Phi$ penetrates the center of the
ring.\label{fig:ring}}
\end{figure}

The proposed method is based on the multichannel scattering technique,
that was developed for studying the tunneling of a single electron in
the presence of scattering by phonons \cite{anda,bonca1}.  Since its
introduction, it has been successfully applied to a variety of
problems where a single electron is coupled to phonon modes
\cite{bonca2,bonca3,ness1,ness2,ness3,torres} and even incorporated into Landauer theory
where the influence of electron-phonon scattering on the
nonequilibrium electron distribution has been investigated
\cite{Emberly}.  We now generalise this method to study many-electron
problems.
While the method can be applied to more general situations, we choose
for simplicity a particular physical system which will also serve as a
toy-model for the study of the e-e interaction induced
dephasing.  We thus consider an AB ring coupled to two ideal
one-dimensional leads, see Fig.~(\ref{fig:ring}).  The ring is
described by a Hubbard-type Hamiltonian
\begin{equation}
\begin{split}
H_\rng = & \sum_{j,\sigma} \left( \epsilon\ \cp{j}{\sigma} \cc{j} {\sigma} 
- t e^{i \phi_j} \cp{j+1}{\sigma} \cc{j}{\sigma} + \textrm{h.c.} \right) \\
& + U \sum_j \cp{j}{\upa} \cc{j}{\upa} \cp{j}{\dna} \cc{j}{\dna}
\end{split}
\label{hring}
\end{equation}
The operator $\cp{j}{\sigma}$ creates an electron with spin $\sigma$
at site $j$ and we make a formal identification 
$\cp{7}{\sigma}=\cp{1}{\sigma}$.
The phases $\phi_j$ describe phase changes due to magnetic
flux penetrating the ring.
The leads are described by the tight-binding Hamiltonian
\begin{eqnarray*}
{H_{\text{lead}}} = -t_{\text{lead}} \sum_{i,\sigma} \cpm{i+1}{\sigma} \cm{i}{\sigma} + \textrm{h.c.} \\
-t_{\text{lead}} \sum_{i,\sigma} \cpn{i+1}{\sigma} \cn{i}{\sigma} + \textrm{h.c.}
\end{eqnarray*}
The operator $\cpm{i}{\sigma}$ creates an electron with spin $\sigma$ at site $i$
on the left lead, while the operator $\cpn{i}{\sigma}$ does the same on the right lead.
The ring is coupled to the electrodes with coupling constants $t_0$:
\begin{equation}
H_c = -t_0 \sum_{\sigma} \left( \cpm{1}{\sigma} \cc{1}{\sigma} + \textrm{h.c.} \right)
-t_0 \sum_{\sigma} \left( \cpn{1}{\sigma} \cc{4}{\sigma} + \textrm{h.c.} \right)
\end{equation}

Our approximation consists of taking into account only
those many-electron states in which at most one (scattering)
electron is located
outside the ring. 
Before the impact of the electron, there are $n=n_\upa+n_\dna$ other
electrons bound on the AB ring. We truncate all many-body states,
where additional electrons hop from the interacting region to the
lead. When  physical parameters of the system, {\it e.g.}
$(\epsilon,t,U)$, are
chosen in such a way that these $n$ electrons are bound in the
interacting region, the approximation is equivalent
to neglecting the exponentially decaying tails of the $n$-electron
wavefunction in the leads. 
Before the scattering, the bound electrons are therefore in one of
the $n$-particle eigenstates of the Hamiltonian $H_\rng$,
Eq.~(\ref{hring}). We denote
these states by $\ket{\alpha_i^\upa}$ and their energies by $\epsilon_i^\upa$.
When an incoming electron with spin up enters the ring, 
the system is in one of the $n+1$-particle states which we 
denote by $\ket{\beta_i}$. These states are not necessarily
eigenstates of $H_\rng$. 
After the scattering there is a single electron in
one of the leads, while the ring is again in one of the $n$-particle
eigenstates of $H_\rng$.
This state can either be one of the $\ket{\alpha_i^\upa}$
states or (if the spin of the scattering electron has been flipped)
in one of the $n$-electron
eigenstates with $n_\upa+1$ spin-up electrons   and
$n_\dna-1$  spin-down electrons. Spin-flipped states are labeled 
by $\ket{\alpha_i^\dna}$ and their energies by $\epsilon_i^\dna$.

By taking into account only the allowed states, the wave-function that
describes the scattering of one electron on the AB ring is given by
\begin{equation}
\begin{split}
\ket{\Psi} = &
\sum_{i=1}^{\infty} \sum_{j,\sigma} d^\lll_{i,j,\sigma} \cpm{i}{\sigma} \ket{\alpha_j^\sigma}
+\sum_{i=1}^{\infty} \sum_{j,\sigma} d^\rr_{i,j,\sigma} \cpn{i}{\sigma} \ket{\alpha_j^\sigma} \\
& + \sum_j e_j \ket{\beta_j}.
\end{split}
\label{psi}
\end{equation}
We consider a steady-state scattering described by the
the Schr\"odinger equation $H\ket{\Psi}=E\ket{\Psi}$ with $H=H_\rng+H_{\text{lead}}+H_c$.
We operate on this equation from the left with $\bra{\beta_l}$ and
using Eq.~(\ref{psi}) we
obtain
\begin{equation}
\label{brezprune}
-t_0 \sum_{j,\sigma} b^\lll_{l,j,\sigma} d^\lll_{1,j,\sigma} 
-t_0 \sum_{j,\sigma} b^\rr_{l,j,\sigma} d^\rr_{1,j,\sigma}
+ \sum_{k} h_{l,k} e_k = E e_l,
\end{equation}
where  $b$'s denote scalar products
$b^\lll_{l,j,\sigma}=\bra{\beta_l} \cp{1}{\sigma}
\ket{\alpha_j^\sigma}$ and $b^\rr_{l,j,\sigma}=\bra{\beta_l}
\cp{4}{\sigma}
\ket{\alpha_j^\sigma}$, while $h_{l,k}=\braket{\beta_l|H_\rng}{\beta_k}$
are the matrix elements of Hamiltonian $H_\rng$ in the $n+1$ electron
subspace.
By operating with $\bra{\alpha_j^\sigma} \cm{1}{\sigma}$ from the left
we get
\begin{equation}
\label{pred}
-t_{\text{lead}} d^\lll_{2,j,\sigma} 
-t_0 \sum_k (b^\lll_{k,j,\sigma})^\ast e_k + 
\epsilon_j^\sigma d^\lll_{1,j,\sigma} = E d^\lll_{1,j,\sigma}.
\end{equation}
In an open outgoing channel $(j,\sigma)$ a plane wave can propagate, 
so that $d^\lll_{2,j,\sigma}=\exp(i k_{j,\sigma}) d^\lll_{1,j,\sigma}$.
By   energy conservation the wave number $k_{j,\sigma}$ is obtained 
from  $\epsilon_0-2t_{\text{lead}} \cos(K)=
\epsilon^\sigma_j-2t_{\text{lead}} \cos(k_{j,\sigma})$.
The energy $\epsilon_0$ is the initial energy
of the $n-$ electron bound state on the ring,
$K$ is the wave number of the incoming electron,
and $\epsilon^\sigma_j$ is the final energy of the bound electrons.
Eq.~(\ref{pred}) can thus be written as 
%
%\begin{equation}
%-t_0 \sum_k (b^\lll_{k,j,\sigma})^\ast e_k = 
%(E-\epsilon_j^\sigma+t_{\text{lead}} \exp(i k_{j,\sigma})) d^\lll_{1,j,\sigma}.
%\label{poprune1}
%\end{equation}
\begin{equation}
d^\lll_{1,j,\sigma}=
{-t_0 \sum_k (b^\lll_{k,j,\sigma})^\ast e_k \over  
(E-\epsilon_j^\sigma+t_{\text{lead}} \exp(i k_{j,\sigma})) }.
\label{poprune1}
\end{equation}
Similar equation can be obtained for exponentially decaying outgoing
channels that we also take into account. These are defined through the
relation $\epsilon_0-2t_{\text{lead}}
\cos(K)=\epsilon^\sigma_j-2t_{\text{lead}} \cosh(k_{j,\sigma})$ and
$d^\lll_{2,j,\sigma}=\exp(-\kappa_{j,\sigma}) d^\lll_{1,j,\sigma}$.
Using this and Eq.~({\ref{poprune1}), the leads can be removed
(pruned)  from the problem \cite{bonca1}.

At zero temperature, the electron scatters on the ground state of the
$n-$ particle state in the ring, $\ket{\alpha_0^\upa}$ with the
energy $\epsilon_0$.
In the incoming channel we have both the incoming and outgoing waves,
$d^\lll_{m,0,\upa}=\exp(-i K m)+r \exp(i K m)$. We obtain
$d^\lll_{2,0,\upa}=\exp(i K) d^\lll_{1,0,\upa}+\exp(i K)-\exp(-i K)$.
The equation for the incoming channel thus contains %%% generates 
an additional inhomogeneous term $\exp(i K)-\exp(-i K)$.

Equations \eqref{brezprune}, \eqref{poprune1} and equivalent
equations for the outgoing channels in the right lead
form a system of equations for unknowns $d^\lll_{1,j,\sigma}$,
$d^\rr_{1,j,\sigma}$ and $e_j$.
This system is solved for different energies of the incoming electron
using the conjugated gradients squared (CGS) method with a Jacobi
preconditioner.
The partial transmittivity through  channel $(j,\sigma)$ is given by
$T_{j,\sigma}(E)=\sin(k_{j,\sigma})/\sin(K) |d^\rr_{1,j,\sigma}|^2$. 
Since the method is based on exact solution of many-electron problem, 
we can compute transmission at arbitrarily large values of $U$.

Results can be improved by extending the interaction region, which is
solved numerically by the Lanczos method, by adding additional sites
from the leads. This procedure increases the computational Hilbert
space, and consequently it properly takes into account decaying tails
of bound electron wavefunctions in the leads.  These improvements
mainly lead to energy shifts of the resonance peaks while the general
characteristics of the spectra remain unchanged. In principle, the
region could be extended until the desired convergence is achieved.
In our calculations the interacting region consisted of the ring and
one additional site from each lead.
In cases where the ground state
of the interaction region was degenerate, we averaged the 
transmittivity spectra over all the degenerate states.
The variational space
taken into account in our calculation was equivalent to a Hubbard model
on 8 sites with no translational symmetry. The system of
equations was solved to an accuracy finer than the linewidth
in the calculated spectra.

We now investigate the effect of the interactions on an electron as it
tunnels through the ring.  The on-site energies are
$\epsilon=-4.5t_{\text{lead}}$, and the overlap integrals are $t=\sqrt{3}t_{\text{lead}}$,
and we set $t_{\text{lead}}=1$.  We have limited the energy of the incident
electron to a half of the bandwidth, {\it e.g.}  $E=[-2,0]$, in order
to avoid ionisation of the bound electron on the ring which would lead
to two electrons exiting the interaction region. In all cases, the
incoming electron was chosen to have spin up.

For a test case we first consider a single electron with spin down
bound on the ring.
We start with the noninteracting case.  In the absence of the magnetic field 
the  transmission reaches unity at the resonance, Fig.~(\ref{fig:1}a).
The electron is fully reflected at any incident energy when the magnetic
flux is $\Phi=\pi$, Fig.~( \ref{fig:1}b). This is the usual
Aharonov-Bohm effect.
\begin{figure}
\includegraphics[angle=-90,clip,totalheight=7cm]{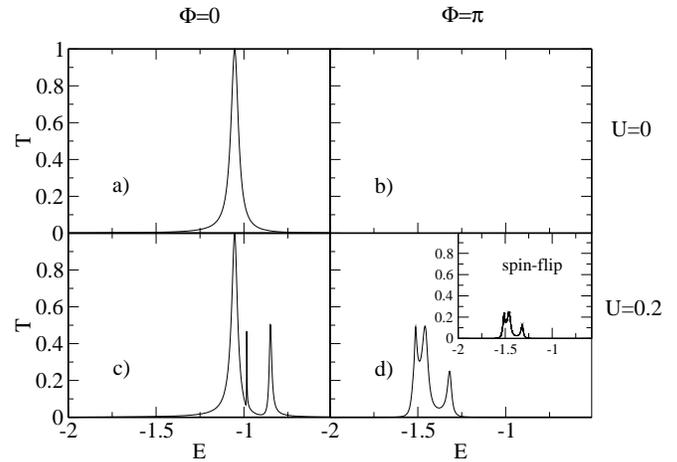}
\caption{Transmission probability as a function of the incident
electron energy for one electron with spin down bound
on the ring. The incoming electron had the spin up. The coupling to
the lead is  $t_0=0.4$. In all cases transmission is purely elastic.
\label{fig:1}}
\end{figure}

We now turn on the interaction. At $\Phi=0$ we still see  a unitary peak 
at the energy of the single-electron resonance, followed by  smaller satellite
peaks caused by the  interaction, Fig.~(\ref{fig:1}c). At $\Phi=\pi$,
when in the absence of $U$ the  electron is fully reflected,
we obtain very high 
transmission probability despite relatively  small $U=0.2$,
Fig.~(\ref{fig:1}d).
In the largest peak the transmission approaches the value  $1/2$.
Since the incoming electron and the bound electron are not entangled,
their total spin is not well defined, therefore 
the total wavefunction is a superposition of a singlet and a triplet
state with equal amplitudes: $\ket{\uparrow \downarrow}=1/\sqrt{2}
\left (\ket{S=1,S_z=0} +  \ket{S=0,S_z=0}\right)$.
The triplet scattering has zero transmission probability at $\Phi=\pi$
since two electrons with spin-up do not interact.
The singlet scattering, however,  reaches the unitary limit at finite
$U$ at the main resonance peak. Averaging over both contributions,
we get $T=1/2$.

The spin-flip scattering part
of the transmission probability is shown in
the inset in Fig. (\ref{fig:1}d). 
The spin-flip and normal scattering contribute equally 
to the total transmission probability. Both are purely 
elastic with respect to energy changes. 

Transmission in the spin-flip channel occurs because
a spin-flipped electron loses its phase memory \cite{ab}.
Transmission without the spin-flip occurs
because of the double degeneracy of the ground state of 
the bound electron. The tunneling
electron changes the symmetry of the bound electron 
wave-function and it thereby acquires a phase shift.
Such symmetry-changing transitions are only possible 
when the electrons interact.

\begin{figure}
\includegraphics[angle=-90,clip,totalheight=7cm]{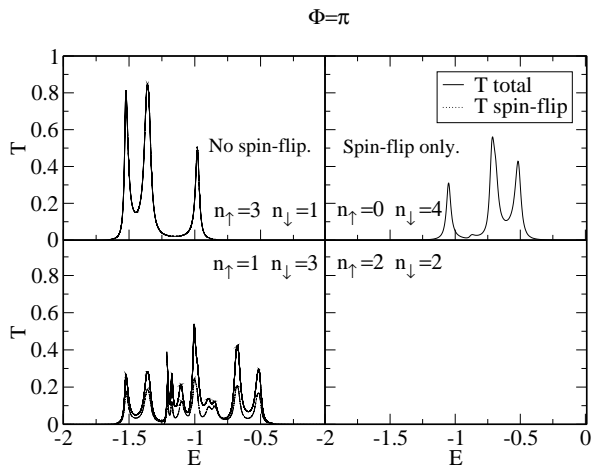}
\caption{Transmission probability as a function of the incident
electron energy for $n_\upa$ ($n_\dna$) electrons with spin up (down).
Incoming electron had  spin up.
Interaction $U=1.0$, coupling constant $t_0=0.3$. \label{fig:2}}
\end{figure}

We now focus on the case of several strongly interacting 
($U=1$) bound electrons when the flux is $\Phi=\pi$.
When the bound state consists of three electrons with spin up
and one electron with spin down ($n_\upa=3, n_\dna=1$,
see Fig.~(\ref{fig:2}))
no spin-flip scattering is possible because such processes
are energetically impossible. The ground state is however
fourfold degenerate and the tunneling electron can get
through the ring at finite $U$ by changing the symmetry of the
many-electron state on the ring. Since the ground state is degenerate,
this process is purely elastic.
In the case of $n_\upa=0, n_\dna=4$ the ground state is nondegenerate, 
however the spin-flip processes are energetically allowed.
We therefore obtain transmission probability only in 
spin-flipped channels. Since in this case the ground state is not
degenerate,
the transmission consists of purely inelastic processes.

In the case when the  ground state is degenerate and
the spin-flips are allowed, we expect dephasing to occur
with or without spin flip. Such is the case of 
$n_\upa=1, n_\dna=3$. The transmittivity without spin-flip 
is purely  elastic, while the spin-flip processes 
are predominantly elastic, with small contribution 
from inelastic channels.
Finally, for $n_\upa=2, n_\dna=2$ electrons are fully reflected from
the ring  since there are no allowed spin-flip nor symmetry-changed 
channels.

\begin{figure}
\includegraphics[clip,totalheight=8.5cm]{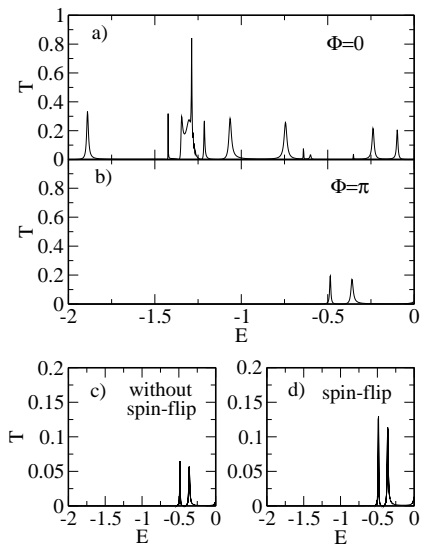}
\caption{Transmission probability as a function of incident
electron energy. for  $n_\upa=2, n_\dna=2$,  $U=15$ and $\epsilon=-20$.
\label{fig:3}}
\end{figure}

Finally we show the influence of large $U=15$ on the case of $n_\upa=2,
n_\dna=2$, where at $U=1$ due to widely spaced many-electron levels
transmission remained zero in the whole interval of incoming electron
energy.  At large $U=15$ the energy difference between the
nondegenerate ground state and the first excited state decreases in
comparison with $U=1$ case, as the states become compressed in the
lower Hubbard band.  We changed the on-site energy to $\epsilon=-20$
in order to keep the electrons bound on the ring.  At $\Phi=0$ there
are several energies at which the electron can resonantly tunnel
through the ring, Fig.~(\ref{fig:3}a).  At $\Phi=\pi$, the electron
can only tunnel inelastically.  The energy difference to the first
excited state in the $n$ electron Hubbard band is approximately
$1.4$. We find indeed that only the electrons that are more than $1.4$
above the bottom of the energy band can tunnel,
Fig.~(\ref{fig:3}b). Such inelastic processes occur both without
(Fig.~\ref{fig:3}c) or with spin-flip (Fig.~\ref{fig:3}d).

Using a simple model and a new numerical method we have extracted the
essential physics of tunneling  through the AB ring. In particular we
have focused on the the role of e-e interactions on dephasing.  While
the proposed method clearly has some limitations (small interacting
regions, inability to describe ionization processes, neglect of
many-body effects in the leads),
it nevertheless gives precise answers to the question: 'What are the main
dephasing mechanisms caused by the e-e interaction?'

A particle can tunnel through AB ring at $\Phi=\pi$ elastically by a)
changing the symmetry of the many-electron state which is possible in
the case of degeneracy or b) by
flipping the spin. Tunneling can also occur in the inelastic channel
by exciting the many-electron state on the ring into an excited state
with or without the spin-flip. Depending on the number of bound
electrons, their total spin, degeneracy of the ground state and
available energy of the incoming electron, the total transmission can
be composed of partial transmissions caused by either one of the
listed processes.

\begin{acknowledgments}
Authors acknowledge the support of the Ministry of Education, Science
and Sport of Slovenia. 
\end{acknowledgments}

\bibliography{ring}

\end{document}